\documentclass[aps,prb,preprint,showpacs,floatfix,amsmath,amssymb,superscriptaddress]{revtex4-1}
\usepackage{graphicx,bm}
\usepackage{epstopdf}
\begin{document}
%\doublespacing
%\onehalfspacing
%\topmargin 0.22cm
%\topmargin 0.45cm
%{\textbf Supplement Material}

\title{Giant tunneling magnetoresistance in Fe$_2$CrSi/Fe$_2$TiSi/Fe$_2$CrSi magnetic tunnel junction}
\author{He Sun}
\affiliation{Key Laboratory of Physics and Technology for Advanced Batteries (Ministry of Education), College of Physics, Jilin University, Changchun 130012, China}
\author{Zhuang Qu}
\affiliation{Department of Physics, Hunan Normal University, Changsha 410081, China}
\author{Liwei Jiang}
\email[Corresponding~author: ]{jlw@jlu.edu.cn}
\author{Yisong Zheng}
\email[Corresponding~author: ]{zhengys@jlu.edu.cn}
\affiliation{Key Laboratory of Physics and Technology for Advanced Batteries (Ministry of Education), College of Physics, Jilin University, Changchun 130012, China}
%\date{\today}

\begin{abstract}
    %{\textbf
    {We propose a theoretical model for an all-Heusler magnetic tunnel junction that uses two Heusler compounds: Fe$_2$CrSi and Fe$_2$TiSi, both of which can be experimentally synthesized. Fe$_2$CrSi is a half-metallic ferromagnet, making it a promising material for efficient spin injection in magnetic random access memories and other spin-dependent devices. While, Fe$_2$TiSi is a nonmagnetic semiconductor that has the same lattice structure and comparable lattice constant with Fe$_2$CrSi, as it can be obtained by substituting the $Y$-site atoms in Fe$_2$CrSi. By using Fe$_2$TiSi as a tunneling barrier sandwiched by two pieces of semi-infinite Fe$_2$CrSi to construct an all-Heusler magnetic tunnel junction, the interface disorder is naturally reduced. Our calculations demonstrate that this magnetic tunnel junction can exhibit a giant tunneling magnetoresistance of up to 10$^{9}$\% and remains robust under finite bias voltage. These characteristics suggest that Fe$_2$CrSi/Fe$_2$TiSi/Fe$_2$CrSi MTJ will be an ideal candidate for future spintronic applications. More importantly, such a device model can keep such a giant tunneling magnetoresistance at and beyond room temperature due to the high Curie temperature of Fe$_2$CrSi.
}

\end{abstract}
    %Spintronics; Giant magnetoresistance; MTJ; Heusler compounds
    \maketitle

    \section{Introduction}
Spintronics is a field aimed at utilizing both the electron's charge and spin degrees of freedom to address significant issues such as Ohmic losses and difficulty in miniaturization that exist in conventional electronic devices\cite{Wolf,Fabian,Maekawa}. Devices based on spintronics have great application prospects because of their high efficiency, low-energy consumption, great stability and high integration to carry, store and process information\cite{Inomata,Bhatti}. Typically, magnetic random access memory (MRAM) is an example of information storage device that utilizes spintronic technology\cite{Inomata,Bhatti}. At present, the core cell of MRAM is the magnetic tunnel junction (MTJ), because its high tunneling magnetoresistance (TMR) ratio is expected to provide the necessary output voltage swing\cite{Nakagome,Ikeda}. Specifically, the basic structure of an MTJ is a sandwich configuration, with ferromagnetic layers (electrodes) placed on opposite side of a nonmagnetic insulating or semiconducting barrier (spacer) layer. The primary requirement for fully commercialized spintronic devices is the design of MTJs that can operate at or above room temperature (RT). To date, the highest reported TMR ratio at RT is 604\% in the CoFeB/MgO/CoFeB MTJ\cite{Ikeda1}, but to achieve a magnetic switch, the TMR ratio needs to be improved to over 1,000\%\cite{Hirohata}.
\par
To achieve a magnetic switch by further increasing the TMR ratios at RT, MTJs require the use of high spin-polarized ferromagnets, such as half-metallic ferromagnets (HMFs)\cite{Groot}, since the performance of MTJs depends on the spin-polarized current induced in the system. HMFs have 100\% spin polarization at the Fermi level, where one of the spin bands is metallic but the other is semiconducting or insulating. Fe-based double perovskites\cite{Kobayashi,Kobayashi1} and Co-based Heusler compounds\cite{Sakuraba,Tezuka,Marukame} are reported to be HMFs with a high Curie temperature, which makes them ideal candidates for ferromagnetic layers. Nevertheless, their half metallicity can be destroyed by atomic disorder stemming from defects and dislocations around the heterojunction, leading to spin fluctuations that reduce the TMR ratio at finite temperature\cite{Hirohata}. Previous studies have demonstrated that interfacial spin fluctuations can be minimized by improving the interfacial matching between the ferromagnetic layer and the nonmagnetic layer\cite{Hirohata}.
\par
Considering that there are more than 1,000 intermetallic compounds belonging to the Heusler family, which share similar crystal structures and exhibit a wide range of tunable properties\cite{Palms,Wollmann}, including the ability to modify magnetism through element substitution, we are confident that an all-Heusler MTJ can serve as a promising device model for enhancing interfacial matching. In detail, Heusler compounds are classified as binary ($X_3Z$), half ($XYZ$), full ($X_2YZ$), and quaternary ($XX'YZ$) compounds, with transition metal elements occupying $X$, $X'$, and $Y$ sites, and main group elements occupying $Z$ sites. The total magnetic moment $M_t$ of Heusler compounds can be estimated using the Slater-Pauling rule\cite{Slater,Pauling}. Namely, the general relationship between $M_t$ and the total number of valence electrons $N_t$ for $XYZ$ ($XX'YZ/X_2YZ/X_3Z$) compounds is $M_t = N_t - 18$ ($24$). Consequently, an $XYZ$ ($XX'YZ/X_2YZ/X_3Z$) compound with $N_t=18$ ($24$) is non-magnetic, while those with different numbers of valence electrons are magnetic. %From Heusler compounds, we choose the HMF as the ferromagnetic layer and then select an insulating or semiconducting material that matches its lattice as a non-magnetic spacer layer.
\par
In the present work we will focus on MTJ using full Heusler Fe$_2$CrSi as a ferromagnetic-layer material. Fe$_2$CrSi is a promising choice for this role because it has the following advantages\cite{Yoshimura,Ko}. For instance, it is an HMF with a relatively low magnetization of 2 $\mu_B$/f.u., enabling a low critical current for spin transfer torque switching\cite{Mangin}. Moreover, Fe$_2$CrSi possesses a suitable Curie temperature of 520 K, which is low enough for thermally assisted recording but still significantly higher than RT. However, the TMR of MTJ constructed by Fe$_2$CrSi and MgO is not high at RT due to the oxidation of Fe$_2$CrSi at the interface\cite{Wang}. To match Fe$_2$CrSi, we select semiconductor Fe$_2$TiSi as the nonmagnetic layer. Both materials share the same L2$_1$-structure, and Fe$_2$TiSi can be obtained by replacing the $Y$-site atom (Cr$\rightarrow$Ti) with Fe$_2$CrSi. This results in a high lattice matching between them. Successful experimental preparations of Fe$_2$CrSi using the melt-spinning method\cite{Luo} and Fe$_2$TiSi via magnetron sputtering\cite{Meinert} suggest that MTJ composed of these two materials hold great potential value for practical applications.
\par
The structure of the paper is as follows: In Sec. II, we explain the computational methods utilized in this study, including the calculation settings and device model. Moving on to Sec. III, we present the spin-dependent electronic transport properties of the MTJ with a stable interface, where we observe a giant TMR ratio of 10$^{9}$\%. Finally, we summarize our findings in Sec. IV.

\section{Computational details}
All first-principles calculations of this work are performed by using the density functional theory (DFT), implemented within the Vienna $ab$ $initio$ simulation package (VASP)\cite{Kresse,Kresse1} with a projected augmented-wave (PAW) basis\cite{Kresse2} and QuantumATK simulation tool\cite{Smidstrup} with a linear combination of atomic orbital (LCAO) PseudoDojo basis. The former is mainly used for the geometry optimizations, where the exchange-correlation potential is treated with the generalized gradient approximation (GGA) in the form of Perdew-Burke-Ernzerhof (PBE)\cite{Perdew} functional. Additionally, 400 eV is used for plane-wave cutoff energy. The geometries are relaxed until all forces are smaller than 0.05 eV/{\AA}, while a $13 \times 13 \times 13$ and $13 \times13 \times 1$ Monkhorst-Pack \textbf{k}-point mesh are used in the calculations for the bulk material and slab respectively. In order to avoid interactions between adjacent slabs, a vacuum of 20 {\AA} thickness is applied along the stacking direction of [001].
\par
The device prototype of Fe$_2$CrSi/Fe$_2$TiSi/Fe$_2$CrSi (001)-MTJ with the interface located at the (001) crystallographic plane, is illustrated in Fig. 1(a). The MTJ device is initially optimized using QuantumATK, and its spin-dependent electronic transport properties are then calculated using the non-equilibrium Green's function (NEGF) approach\cite{NEGF} within the same software. The exchange-correlation potential we used is GGA-PBE. Cutoff energy of 100 hartree is adopted with Monkhorst-Pack \textbf{k}-point mesh of $13 \times 13 \times 169$ for the self-consistent field calculations. The $I$-$V$ characteristics are calculated within the Landauer-B\"{u}ttiker approach\cite{Landauer}:
\begin{equation}
  I_{p/ap}(V)=\frac{e}{h}\sum_\sigma\int_{-\infty}^\infty{T^\sigma_{p/ap}(E,V)\left[f(E,\mu_L)-f(E,\mu_R)\right]}dE.
\end{equation}
Here, $V=(\mu_L-\mu_R)/e$ denotes the applied bias voltage, with $\mu_L$ and $\mu_R$ being the chemical potentials of the left and
right electrodes respectively. $T^\sigma_{p/ap}(E,V)$ is a spin($\sigma$)-dependent transmission coefficient at a certain electron energy $E$ and a certain bias voltage $V$ in parallel/anti-parallel magnetization configuration. $f(E,\mu_L/\mu_R)$ is the Fermi-Dirac distribution of the left/right electrode. For the calculation of $T^\sigma_{p/ap}(E,V)$, we choose a dense $40\times40$ $\textbf{k}_\parallel$-point mesh. To be precise, the transmission coefficients in Eq. (1) are inherently dependent on the bias magnitude. Hence, when considering an arbitrary bias, it is necessary to calculate them self-consistently. However, as an alternative approach, we can neglect this bias dependence by substituting the numerical results of the transmission coefficients under zero bias into Eq. (1) for computing the $I$-$V$ curve. This straightforward technique, known as the linear response approximation, has frequently been utilized in earlier researches.

\section{Results and discussion}
\begin{figure}
 \centering
 \includegraphics[width=15.1cm]{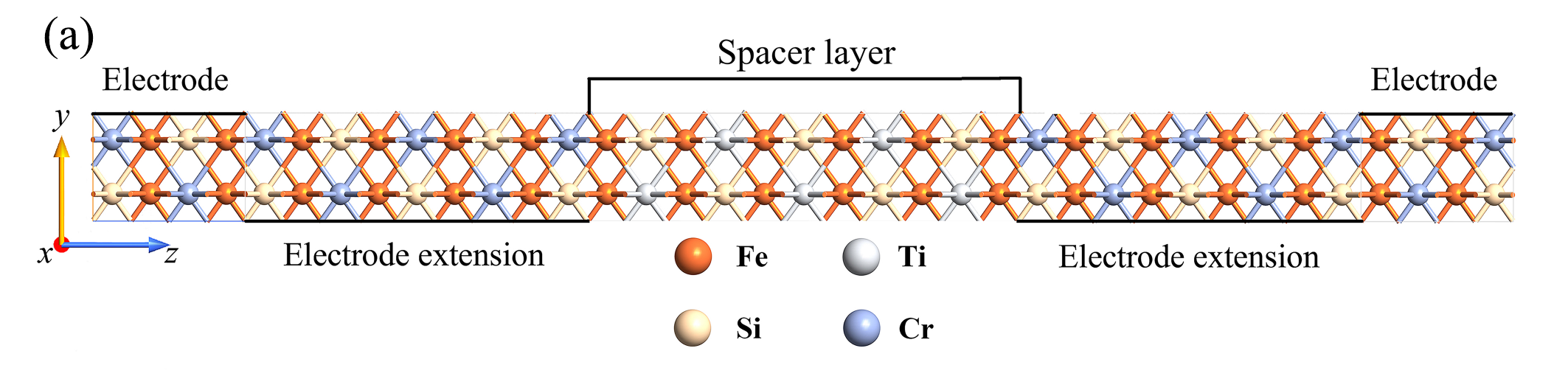}
 \includegraphics[width=15.0cm]{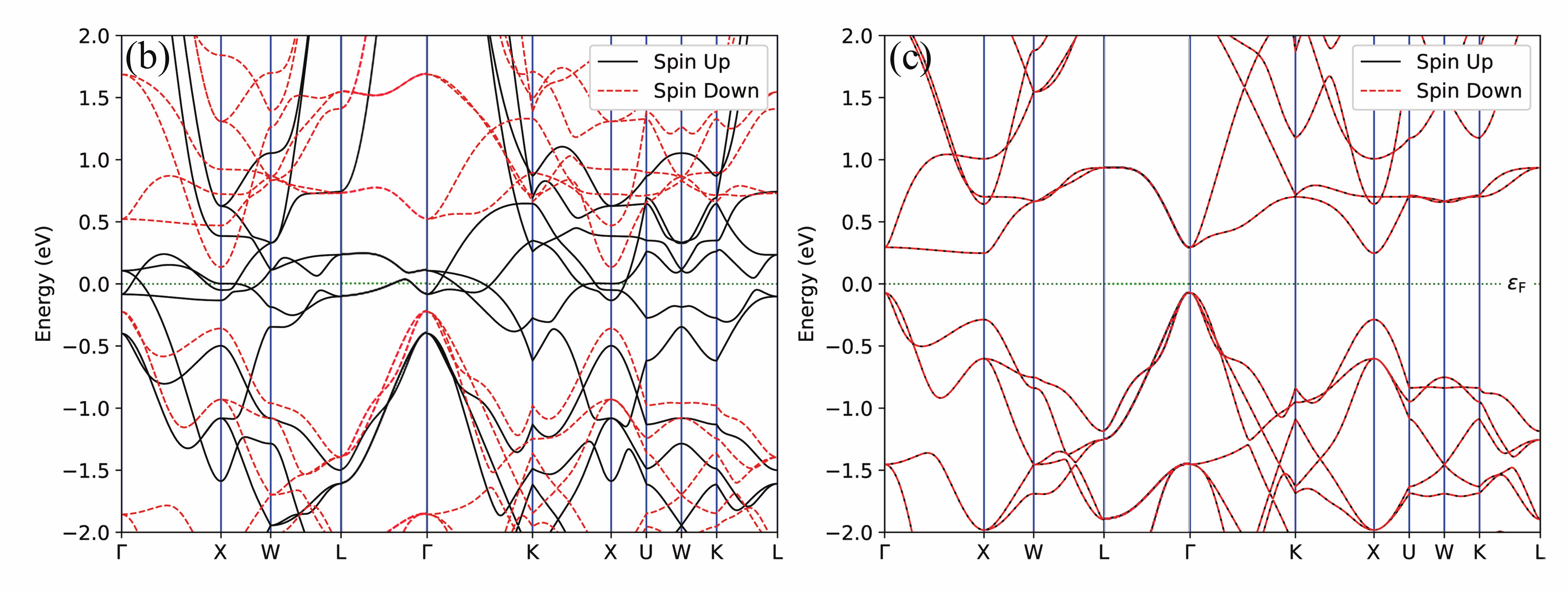}
 \caption{(Color online) (a) The atomic structure of the Fe$_2$CrSi/Fe$_2$TiSi/Fe$_2$CrSi tunnel junction with the electronic transport direction in $z$ direction. The device is consisted of two (left and right) semi-infinite electrodes, electrode extensions and the spacer layer in between. Band structure of Fe$_2$CrSi (b) and Fe$_2$TiSi (c). The dashed green line denotes the Fermi level that is set to zero.}\label{schematic}
\end{figure}
\par
Both Fe$_2$CrSi and Fe$_2$TiSi are full-Heusler compounds with a cubic space group {$Fm\overline{3}m$} (no. 225) in {$L2_1$} structure. The calculated lattice constants of Fe$_2$CrSi and Fe$_2$TiSi are 5.59 {\AA} and 5.65 {\AA}, respectively, so there is only $\sim1{\%}$ lattice mismatch between them. Such a mismatch is sufficiently small to avoid interface disorder. Then the calculated band structures of these two Heusler compounds are illustrated in Figs. 1(b) and (c). The spin-up band of Fe$_2$CrSi exhibits metallic behavior, crossing the Fermi level, while the spin-down band displays an indirect band gap of 0.35 eV. This results in a complete spin polarization of the conduction electrons at Fermi level and makes Fe$_2$CrSi half metallic. Since the Fermi level is situated approximately at the midpoint of the gap, half metallicity remains stable against external influences such as thermal excitation to smear the high spin polarization of a half metal. Furthermore, when Fe$_2$CrSi is grown as a thin film on a substrate, its lattice constant can be influenced by the substrate lattice. However, its half metallicity has been reported to remain robust even when the lattice constant of Fe$_2$CrSi varies between 5.51 and 5.73 {\AA}\cite{Luo}. In a word, Fe$_2$CrSi is well-suited as a ferromagnetic layer in MTJ and shows potential for use in spintronic devices operating at RT. Regarding Fe$_2$TiSi, it is a non-magnetic semiconductor that follows the Slater-Pauling rule\cite{Slater,Pauling}, with a calculated indirect band gap of 0.32 eV. The lattice and band matching between Fe$_2$CrSi and Fe$_2$TiSi suggest that they are excellent candidate materials for use in MTJs.
\begin{figure}
  \centering
  \includegraphics[width=14cm]{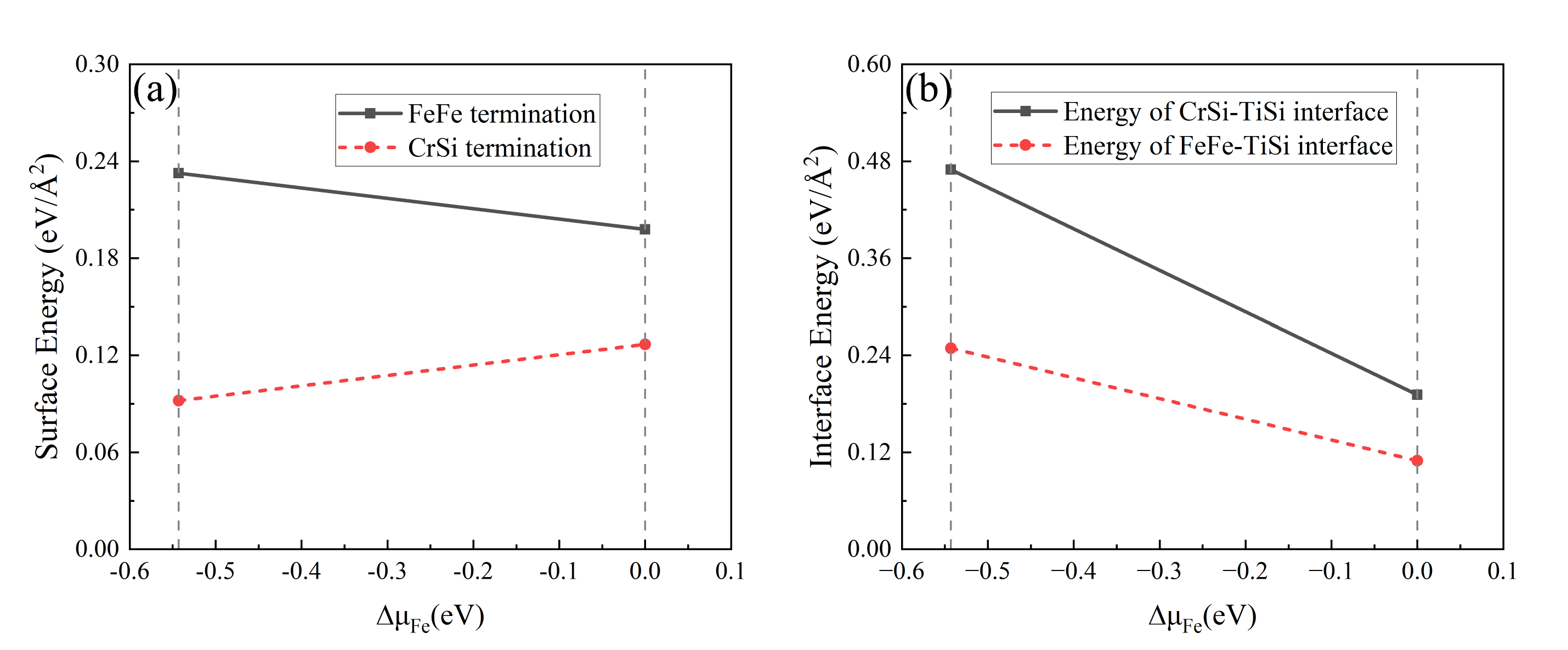}
  \caption{(Color online) (a) The surface energies of FeFe- and CrSi- terminations. (b) The formation energies of CrSi-FeFe and CrSi-TiSi interfaces.}\label{fig1}
\end{figure}
\par
Before constructing the MTJ device, we need to determine the most stable interface along the [001] direction. The growth sequence of the multilayer film in the MTJ device starts with the left electrode material, passes through the non-magnetic barrier layer, and ends with the right electrode. Therefore, our first step is to determine the stable surface termination of the electrode Fe$_2$CrSi where a few atomic layers of Fe$_2$TiSi will be grown. We consider two symmetric slabs of Fe$_2$CrSi with the same thickness (13 atomic layers) but different surface terminations, i.e. FeFe- or CrSi-termination. Using the method of $ab$ $initio$ atomistic thermodynamics\cite{Weinert,Scheffler}, we can calculate the surface energy $\gamma_{S}=\left[ E^{slab}-\sum_i{N_i\mu _i} \right]/{2A}= \left[ \Delta E_{f}^{slab}-\sum_i{N_i\Delta \mu _i} \right]/{2A}$. Here, $A$ denotes the surface area; $E^{slab}$ is the total energy of the symmetric slab.
$N_i$ counts the number of atom $i$($i=X,Y,$ or $Z$), and $\mu _i$ is the chemical potential of atom $i$ in bulk $X_2YZ$. $\Delta E_{f}^{slab}$ represents the formation energy of the slab, given by $\Delta E_{f}^{slab} = E^{slab} - \sum\nolimits_i{N_i\mu_{i}^{bulk}}$, where $\mu_{i}^{bulk}$ denotes the chemical potential of $i$ in its most stable elemental solid phase. The relative chemical potential is denoted as $\Delta\mu_i=\mu_i-\mu_i^{bulk}$, and only when $\Delta\mu_i<0$ in the growing environment, the surface of the compound can be grown, otherwise it facilitates the formation of the elemental solid phase of atom $i$. Additionally, due to the chemical equilibrium between the surface phase and the underlying bulk phase in the slab, we have $\sum_i \Delta\mu_i=\Delta E^{X_2YZ}$, where $\Delta E^{X_2YZ}$ represents the formation energy of the bulk phase of $X_2YZ$. Therefore, for Fe$_2$CrSi, $2\Delta\mu_\text{Fe}=\Delta E^{\text{Fe}_2\text{Cr}\text{Si}}-\Delta\mu_\text{Cr}-\Delta\mu_\text{Si}$, providing a lower limit for $\Delta\mu_\text{Fe}$ since $\Delta\mu_\text{Cr/Si}<0$. The numerical result can be seen in Fig. 2(a), showing that the CrSi-termination is more stable than the FeFe-termination.
\par
Next, we can determine the stable interface type for growing Fe$_2$TiSi on the CrSi-terminated surface of Fe$_2$CrSi based on the formation energy of the interface. When the thin barrier material is grown on the left electrode, the in-plane lattice constant of the barrier material is imposed. In this case, we construct two types of Fe$_2$CrSi/Fe$_2$TiSi/Fe$_2$CrSi slabs with structural symmetry but different interfaces: CrSi-FeFe or CrSi-TiSi. The in-plane lattice constant is fixed by Fe$_2$CrSi, while the out-of-plane lattice constant is allowed to relax freely. Using the method of $ab$ $initio$ atomistic thermodynamics, the interface energy is defined as $\gamma^{IF}= \left[E^{IF}-\sum_i{N_i \mu _i}-2A \gamma_{S} \right]/{2A}$. Here, $E^{IF}$ is the total energy of the composite slab with one specific interface type, and $\gamma_{S}$ is the CrSi-surface energy of Fe$_2$CrSi (calculated earlier). Following the preceding argument, the interface energy can be expressed as a function of $\Delta\mu_\text{Fe}$, and the numerical results are shown in Fig. 2(b). We can conclude that the formation energy of the CrSi-FeFe interface is smaller than that of the CrSi-TiSi interface. This suggests a preference for the interface to adopt a normal atomic order, wherein the MTJ structure maintains the same order of atomic planes as the bulk Fe$_2$CrSi, with the only distinction being the substitution of Cr for Ti in the barrier layer at the atomic level, as shown in Fig. 1(a). Therefore, we refer to such a kind of MTJ as the MTJ with normal atomic order. Previous reports have indicated that such a normal atomic order at the interface of an all-Heusler device can preserve the half metallicity of the $L2_1$-structure Heusler compound\cite{Fesler,interface}.
\begin{figure}
\centering
  \includegraphics[width=17cm]{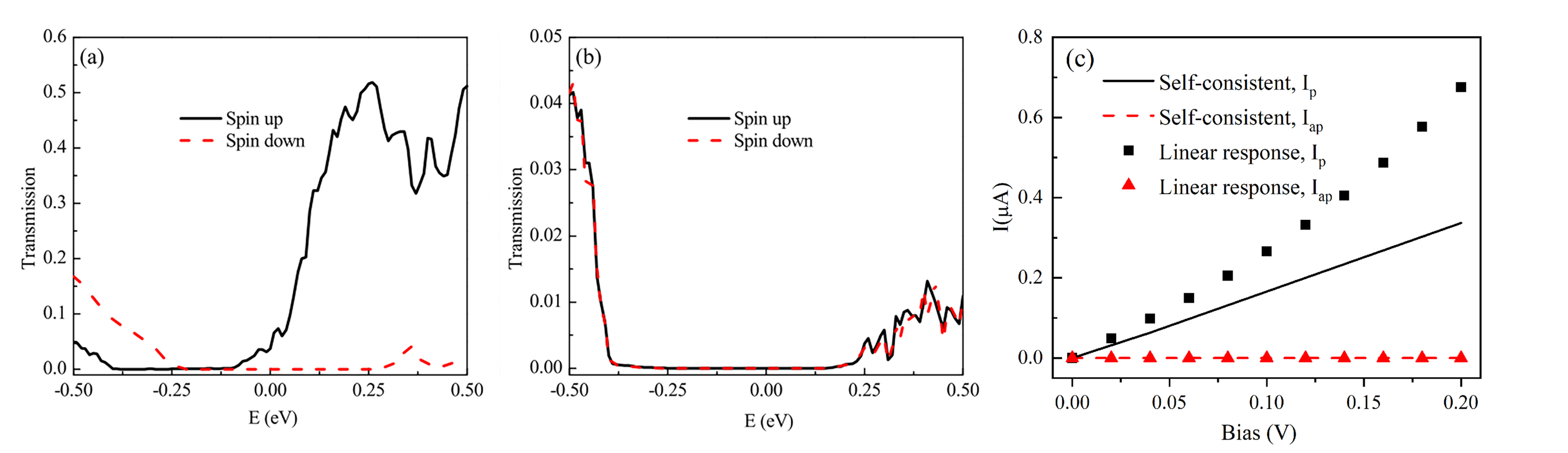}
  \caption{(Color online) The spin-resolved transmission spectra of the Fe$_2$CrSi/Fe$_2$TiSi/Fe$_2$CrSi MTJ in (a) parallel and (b) anti-parallel magnetization configuration, respectively. (c) The $I$-$V$ curves of Fe$_2$CrSi/Fe$_2$TiSi/Fe$_2$CrSi MTJ as a function of bias voltage.}\label{fig2}
\end{figure}
\par
Figure 1(a) shows the Fe$_2$CrSi/Fe$_2$TiSi/Fe$_2$CrSi MTJ with the most stable interface, in which an 11 atomic layer (1.4 nm) of semiconductor Fe$_2$TiSi is sandwiched between semi-infinite left and right Fe$_2$CrSi magnetic electrodes. In the following we focus on the spin-dependent electronic transport properties of such an MTJ. The zero bias transmission spectra of MTJ are presented in Figs. 3(a) and (b). In the case when the two electrodes are magnetized in parallel, as depicted in Fig. 3(a), there is a substantial spin-up transmission coefficient of 0.04 and nearly zero spin-down transmission coefficient at the Fermi level. The conductance is equal to the transmission coefficient multiplied by $e^2/h$, namely, the spin-up conductance is 0.04 $e^2/h$ when the MTJ is in parallel magnetization configuration. This value is an order of magnitude larger than the corresponding conductance of another Fe$_2$TiSi-based MTJ (CoFeTiSi/Fe$_2$TiSi/CoFeTiSi) mentioned in the recent literature\cite{Feng}. As shown in Fig. 3(b), when the MTJ is in the anti-parallel configuration, the transmission coefficients of spin-up and spin-down almost coincide with each other due to the structural symmetry of the MTJ, and both are close to zero around the Fermi level. The reason for the significantly higher spin-up transmission coefficient in parallel magnetization configuration compared to other cases can be easily understood by examining the band structure shown in Figs. 1(b) and (c). In parallel magnetization configuration, around the Fermi level the left and right electrodes Fe$_2$CrSi can only provide the spin-up electronic states. As a result, the spin-up electrons incident from the left electrode can tunnel through the barrier and occupy the spin-up state on the right. This leads to a larger transmission coefficient. In contrast, the nearly zero transmission coefficient for spin-down electrons in the parallel magnetization configuration, as well as the transmission coefficients for spin-up and spin-down electrons in the anti-parallel magnetization configuration, can be ascribed to the absence of electronic states in both electrodes, or in either the left or right electrode.
\begin{figure}
  \centering
  \includegraphics[trim=0 12cm 0 2cm, clip,width=17cm]{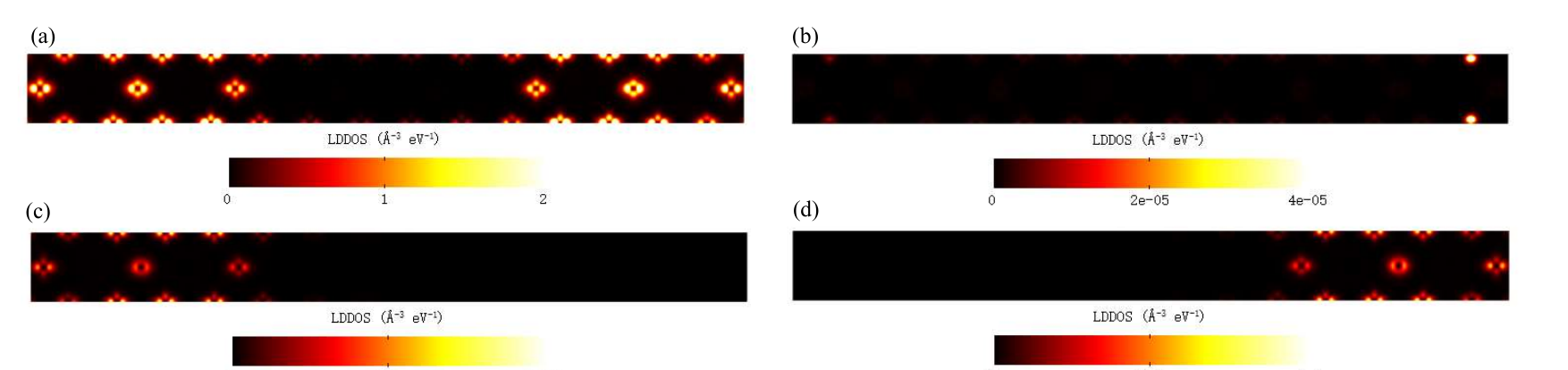}
  \caption{(Color online) The spin-resolved LDDOS of Fe$_2$CrSi/Fe$_2$TiSi/Fe$_2$CrSi MTJ. (a) and (b) The LDDOS of spin-up and spin-down channels in parallel magnetization configuration, respectively. (c) and (d) The LDDOS of spin-up and spin-down channels in anti-parallel magnetization configuration, respectively. }\label{fig3}%In (a), the bright positions in the middle row are Cr atoms with the subscript `1' labeling the row number as shown in Fig. 1(a), and the dark positions on the right of Cr$_1$ are Fe$_1$ atoms. The bright positions in the bottom row of atoms are Fe$_2$ atoms, and the dark positions on the left of Fe$_2$ are Si$_2$ atoms. At last, the top row is a repetition of the bottom row.
\end{figure}

\par
Furthermore, the transmission spectra can be explained by examining the spin-resolved local device density of states (LDDOS) at the Fermi level. Fig. 4 presents the LDDOS covering both electrode extensions and the tunnelling barrier in between. In the case of parallel magnetization configuration, as illustrated in Fig. 4(a), the left and right electrodes provide spin-up electronic states, and the barrier layer Fe$_2$TiSi provides the channels for spin-up electron tunneling. For the spin-down case in Fig. 4(b), both the left and right electrodes have few DOS, which cannot provide spin-down electrons for transport. Thus the transmission coefficient is near zero. In the case of anti-parallel magnetization configuration, the spin-up
and spin-down channels are blocked either in the right(Fig. 4(c)) or left(Fig. 4(d)) electrode, resulting in near zero transmission coefficient at the Fermi level. Due to the electronic transverse wavevector $\textbf{k}_\parallel=(k_x,k_y)$ remains a conserved quantity in the MTJ model. The electronic transmission coefficient is a sum of the ones in individual $\textbf{k}_\parallel$-subspace. Namely, $T^\sigma_{p/ap}(E)=\sum_{\textbf{k}_\parallel} T^\sigma_{p/ap}(E, \textbf{k}_\parallel)$. Then in Fig. 5 we show the $\textbf{k}_\parallel$-resolved transmission coefficients to further clarify the momentum information. As shown in Fig. 5(a), for the spin-up case when the MTJ is in parallel magnetization configuration, the transmission is mainly concentrated at four vertices of the $\textbf{k}_\parallel$-subspace, followed by the region around the $\textbf{k}_\parallel$ = (0, 0) point. This demonstrates strong resonance characteristics, resulting in a large transmission coefficient. In the remaining cases, specifically the spin-down case in parallel magnetization configuration(Fig. 5(b)), and the spin-up (Fig. 5(c)) and spin-down (Fig. 5(d)) cases in anti-parallel magnetization configuration, the transmission coefficient is primarily contributed by the points around $\textbf{k}_\parallel$ = (0, 0), albeit with a noticeably lower intensity compared to the spin-up one in parallel magnetization configuration.
\begin{figure}
  \centering
  \includegraphics[width= 15cm]{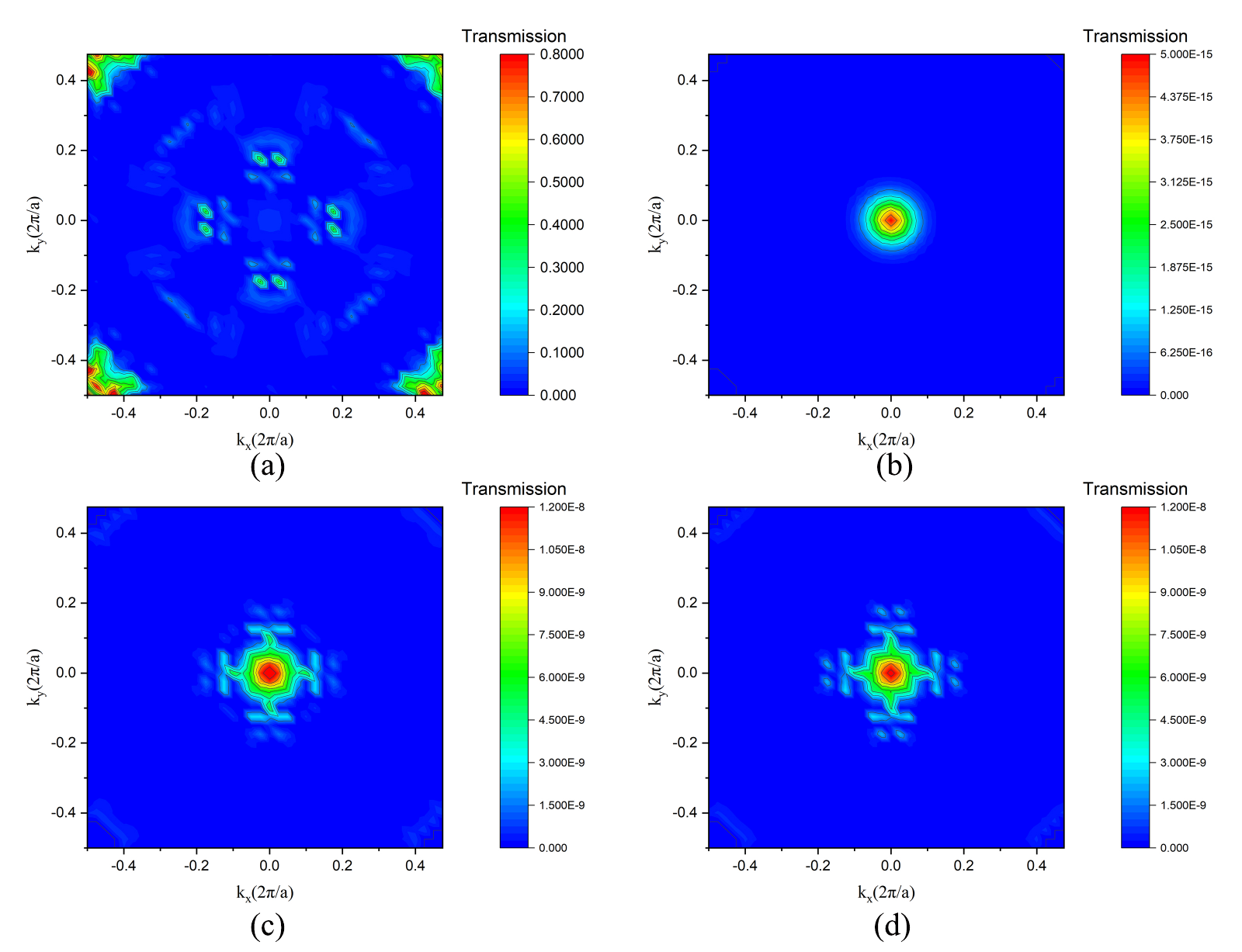}
  \caption{(Color online) The $\textbf{k}_\parallel$-resolved transmission coefficients of Fe$_2$CrSi/Fe$_2$TiSi/Fe$_2$CrSi MTJ for (a)(c) parallel and (b)(d) anti-parallel magnetization configurations at Fermi level. (a) and (c) indicate the result of spin-up. (b) and (d) indicate the result of spin-down.}\label{fig4}
\end{figure}
\par
In order to measure the performance of MTJs, the TMR ratio is a significant parameter, thus we calculate it from $\text{TMR}=\left(G_{p}-G_{ap} \right)/G_{ap}\times 100\%$. Here, $G_{p}=e^2/h[T^{\uparrow}_{p}(E)+T^{\downarrow}_{p}(E)]$ and $G_{ap}=e^2/h[T^{\uparrow}_{ap}(E)+T^{\downarrow}_{ap}(E)]$ represent the conductance of the parallel and anti-parallel magnetization configuration, respectively. $\uparrow$ and $\downarrow$ refer to the up and down spins. At present, we are able to attain a giant TMR ratio of up to $10^{9}$\%. MTJs will be subjected to bias in practical applications, which will affect the performance and stability of MTJs. Therefore, it is necessary to consider the influence of bias voltage on TMR ratio at RT. Under a finite bias, the TMR ratio can be calculated by $\text{TMR}=\left( I_{p}-I_{ap} \right)/I_{ap}\times 100\%$, where $I_p$ and $I_{ap}$ are the current of the parallel and anti-parallel magnetization configuration, respectively. Then the $I$-$V$ curves at RT calculated by self-consistent field and linear response approaches are provided in Fig. 3(c). As seen for anti-parallel magnetization configuration, both approaches show the off state of MTJ under finite bias, i.e., $I_{ap}$ is basically zero. While for parallel magnetization configuration, $I_p$ increases as the bias voltage increases. However, the linear response approach always overestimates the current, and at 0.2 V, the linear response result is about twice of the self-consistent field one. Therefore, the linear response approach can only qualitatively calculate the current. Nevertheless, the TMR ratio obtained by these two approaches show us the same order of 10$^{5}$\%, which can certainly achieve a magnetic switch effect\cite{Hirohata}.

\section{Conclusions}
In summary, we have performed first-principles calculations to analyze the spin-polarized electronic transport properties of the Fe$_2$CrSi/Fe$_2$TiSi/Fe$_2$CrSi MTJ. Based on our calculations of surface and interface energies, it is evident that the device configuration with the normal atomic order is energetically stable. This indicates that the barrier layer of MTJ can be obtained simply by substituting Ti atoms for Cr atoms in the central layer of Fe$_2$CrSi along the transport direction. The normal atomic order can effectively reduce the disorder of the interface and minimize the interfacial spin fluctuations, thus maintaining the effective spin-polarized injection of the half-metal electrode\cite{Hirohata,Fesler,interface}. By utilizing the MTJ with normal atomic order, we achieved a high zero-bias TMR ratio of up to 10$^{9}$\%. Furthermore, the Fe$_2$CrSi electrode possesses a suitable Curie temperature and exhibits stable half-metallicity against thermal excitation. As a result, even under finite bias and RT, a considerable TMR ratio of 10$^{5}$\% can be obtained. Based on these findings, we consider the Fe$_2$CrSi/Fe$_2$TiSi/Fe$_2$CrSi MTJ as a promising device model capable of working above RT.

\begin{acknowledgments}
This work was financially supported by the National Natural Science Foundation of China (Grant No. 12174145 and 11774123) and the Fundamental Research Funds for the Central Universities. We thank the High Performance Computing Center of Jilin University for their calculation resource.
\end{acknowledgments}

%\bibliographystyle{apsrev}
%\bibliography{reference}

\begin{thebibliography}{99}

\bibitem{Wolf}S. A. Wolf, D. D. Awschalom, R. A. Buhrman, \emph{et al}., Science \textbf{294}, 1488 (2001).
\bibitem{Fabian}I. \v{Z}uti\'{c}, J. Fabian and S. Das Sarma, Rev. Mod. Phys. \textbf{76}, 323 (2004).
\bibitem{Maekawa}S. Maekawa, S. O. Valenzuela, E. Saitoh and T. Kimura, \emph{Spin Current} (Oxford University Press, Oxford, U.K, 2012).
\bibitem{Inomata}K. Inomata, N. Ikeda, N. Tezuka, R. Goto, S. Sugimoto, M. Wojcik and E. Jedryka, Sci. Technol. Adv. Mater. \textbf{9}, 014101 (2008).
\bibitem{Bhatti}S. Bhatti, R. Sbiaa, A. Hirohata, \emph{et al}., Mater. Today \textbf{9}, 530-548 (2017).
\bibitem{Nakagome}Y. Nakagome, M. Horiguchi, T. Kawahara, and K. Itoh, IBM J. Res. Develop. \textbf{47}, 525 (2003).
\bibitem{Ikeda}S. Ikeda, J. Hayakawa, Y. M. Lee, \emph{et al}., "Magnetic Tunnel Junctions for Spintronic Memories and Beyond," in IEEE Transactions on Electron Devices \textbf{54}, pp. 991-1002 (2007).
\bibitem{Ikeda1}S. Ikeda, J. Hayakawa, Y. Ashizawa, \emph{et al}., Appl. Phys. Lett. \textbf{93}, 082508 (2008).
\bibitem{Hirohata}A. Hirohata, K. Elphick, D. C. Lloyd and S. Mizukami, Front. Phys. \textbf{10}, (2022).
\bibitem{Groot}R. A. de Groot, F. M. Mueller, P. G. van Engen and K. H. J. Buschow, Phys. Rev. Lett. \textbf{50}, 2024 (1983).
\bibitem{Kobayashi}K.I. Kobayashi, T. Kimura, H. Sawada, \emph{et al}., Nature (London) \textbf{395},  677 (1998).
\bibitem{Kobayashi1}K.I. Kobayashi, T. Kimura, Y. Tomioka, \emph{et al}., Phys. Rev. B \textbf{59}, 11159 (1999).
\bibitem{Sakuraba}Y. Sakuraba, T. Miyakoshi, M. Oogane, \emph{et al}., Appl. Phys. Lett. \textbf{89}, 052508 (2006).
\bibitem{Tezuka}N. Tezuka, N. Ikeda, S. Sugimoto and K. Inomata, Appl. Phys. Lett. \textbf{89}, 252508 (2006).
\bibitem{Marukame}T. Marukame, T. Ishikawa, K. Matsuda, \emph{et al}., Appl. Phys. Lett. \textbf{90}, 012508 (2007).
\bibitem{Palms}C. J. Palmstr{\o}m, Prog. Cryst. Growth Charact. Mater. \textbf{62}, 371 (2016).
\bibitem{Wollmann}L. Wollmann, A. K. Nayak, S. S. Parkin, and C. Felser, Annu. Rev. Mater. Res. \textbf{47}, 247 (2017).
\bibitem{Slater}J. C. Slater, Phys. Rev. \textbf{49}, 931 (1936).
\bibitem{Pauling}L. Pauling, Phys. Rev. \textbf{54}, 899 (1938).
\bibitem{Yoshimura}S. Yoshimura, H. Asano, Y. Nakamura, \emph{et al}., J. Appl. Phys. \textbf{103}, 07D716 (2008).
\bibitem{Ko}V. Ko, J. Qiu, P. Luo, \emph{et al}., J. Appl. Phys. \textbf{109}, 07B103 (2011).
\bibitem{Mangin}S. Mangin, Y. Henry, D. Ravelosona, \emph{et al}., Appl. Phys. Lett. \textbf{94}, 012502 (2009).
\bibitem{Wang}Y. -P. Wang, J-J Qiu, H. Lu, Hui,  \emph{et al}., 2013 IEEE 5th International Nanoelectronics Conference (INEC), Singapore, pp. 215-218 (2013).
\bibitem{Luo}H. Luo, Z. Zhu, L. Ma, \emph{et al}., J. Phys. D: Appl. Phys. \textbf{40} 7121-7127 (2007).
\bibitem{Meinert}M. Meinert, M. P. Geisler, J. Schmalhorst,\emph{et al}., Phys. Rev. B \textbf{90}, 085127 (2014).
\bibitem{Kresse}G. Kresse and J. Hafner, Phys. Rev. B \textbf{49}, 14251 (1994).
\bibitem{Kresse1}G. Kresse and J. Furthmu\"{u}ller, J. Computat. Mater. Sci. \textbf{6}, 15 (1996).
\bibitem{Kresse2}G. Kresse and D. Joubert, Phys. Rev. B \textbf{59}, 1758 (1999).
\bibitem{Smidstrup}S. Smidstrup, T. Markussen, P. Vancraeyveld, \emph{et al}., J. Phys.: Condens. Matter \textbf{32}, 015901 (2020).
\bibitem{Perdew}J. P. Perdew, K. Burke, and M. Ernzerhof, Phys. Rev. Lett. \textbf{77}, 3865 (1996).
\bibitem{NEGF}M. Brandbyge, J.-L. Mozos, P. Ordej\'on, J. Taylor, and K. Stokbro, Phys. Rev. B \textbf{65}, 165401 (2002).
\bibitem{Landauer} M. B\"{u}ttiker, Y. Imry, R. Landauer, and S. Pinhas, Phys. Rev. B \textbf{31}, 6207 (1985).

\bibitem{Weinert} C.W. Weinert and M. Scheffler, Phys. Rev. Lett. \textbf{58}, 1456 (1987).
\bibitem{Scheffler} M. Scheffler and J. Dabrowski, Philos. Mag. A \textbf{58}, 107 (1988).
%\bibitem{Zhanglei}L. Zhang, B. Zhang, L. Jiang and Y. Zheng, J. Phys.: Condens. Matter \textbf{34}, 204003 (2022).
\bibitem{Fesler}S. Chadov, T. Graf, K. Chadova, X. Dai, F. Casper, G. H. Fecher, and C. Felser, Phys. Rev. Lett. \textbf{107}, 047202 (2011).
\bibitem{interface}I. Di Marco, A. Held, S. Keshavarz, Y. O. Kvashnin, and L. Chioncel, Phys. Rev. B \textbf{97}, 035105 (2018).
\bibitem{Feng}Y. Feng , H. Ding , X. Li, \emph{et al}., J. Appl. Phys. \textbf{131}, 133901 (2022).

  \end{thebibliography}

\clearpage

\end{document}